\documentclass{ametsocV6.1_noln}
\usepackage{amsmath}
\usepackage{amssymb}

\newcommand{\savg}[1]{\ensuremath{\left\langle #1 \right\rangle}}

\title{How to derive skill from the Fractions Skill Score}

\authors{Bobby Antonio\aff{a,b} and \correspondingauthor{bobby.antonio@physics.ox.ac.uk} 
Laurence Aitchison\aff{c}
}

\affiliation{\aff{a}{Department of Physics, University of Oxford, UK}\\
\aff{b}{School of Geographical Sciences, University of Bristol, Bristol, UK}\\
\aff{c}{Machine Learning and Computational Neuroscience Unit, University of Bristol, Bristol, UK}
}

\abstract{
The Fractions Skill Score (FSS) is a widely used metric for assessing forecast skill, with applications ranging from precipitation to volcanic ash forecasts. By evaluating the fraction of grid squares exceeding a threshold in a neighbourhood, the intuition is that it can avoid the pitfalls of pixel-wise comparisons and identify length scales at which a forecast has skill. The FSS is typically interpreted relative to a `useful' criterion, where a forecast is considered skillful if its score exceeds a simple reference score. However, the typical reference score used is problematic, as it is not derived in a way that provides obvious meaning or that scales with neighbourhood size, and forecasts that do not exceed it can have considerable skill. We therefore provide a new method to determine forecast skill from the FSS, by deriving an expression for the FSS achieved by a random forecast, which provides a more robust and meaningful reference score to compare with. Through illustrative examples we show that this new method considerably changes the length scales at which a forecast would be regarded as skillful, and reveals subtleties in how the FSS should be interpreted. }

\begin{document}

\maketitle

\statement
Forecast verification metrics are crucial to assess accuracy and identify where forecasts can be improved. In this work we investigate a popular verification metric, the Fractions Skill Score, and derive a more robust method to decide if a forecast has sufficiently high skill. This new method significantly improves the quality of insights that can be drawn from this score.

%
\section{Introduction}

Assessing the performance of numerical weather prediction models is crucial for monitoring and guiding model development, and is also extremely challenging, particularly for fields like precipitation that exhibit high spatial variability. One approach to address the double penalty issue that occurs for pixel-wise comparisons \citep{wilks_forecast_2019} is to use aggregated quantities in a neighbourhood around each grid point to assess the change in skill as the neighbourhood size increases~\citep{ebert_fuzzy_2008}. A commonly used score in this category is the Fractions Skill Score (FSS)~\citep{roberts_scale-selective_2008, roberts_assessing_2008}, which evaluates the fractions of grid squares above a certain threshold in a neighbourhood surrounding each grid point. This score has been used to evaluate cutting edge machine learning weather prediction systems~\citep{ayzel_rainnet_2020, ravuri_skilful_2021}, convection permitting models~\citep{woodhams_what_2018, weusthoff_assessing_2010, cafaro_convection-permitting_2021, schwartz_medium-range_2019}, volcanic ash forecasts~\citep{harvey_spatial_2016}, oil spill forecasts~\citep{simecek-beatty_oil_2021}, flood inundation forecasts~\citep{hooker_spatial_2022}, as a loss function for training models~\citep{ebert-uphoff_cira_2021, lagerquist_using_2021, lagerquist_can_2022, price_increasing_2022}, and has been proposed as a replacement for the equitable threat score in operational forecast verification \citep{mittermaier_long-term_2013}.

The Fractions Skill Score is typically categorised as a `neighbourhood' approach to forecast verification \citep{ebert_fuzzy_2008, gilleland_intercomparison_2009} for which the quality of forecasts are measured by comparing the neighbourhoods around each grid square. The result of aggregating features over neighbourhoods has the effect of blurring the forecast and observations, and so in the original FSS formulation was introduced as a way to mitigate the double penalty problem and assess the length scale at which the forecast becomes of high enough quality. Alternatively, the use of neighbourhoods can be interpreted as a way to resample the probability distribution of forecasts and observations \citep{theis_probabilistic_2005}, which then motivates the use of probabilistic scores such as the Brier Skill Score \citep{brier_verification_1950} and Brier Divergence Skill Score \citep{stein_evaluation_2024}. Other neighbourhood approaches include comparing ordered samples within neighbourhoods \citep{rezacova_radar-based_2007}, upscaling the data before comparison \citep{marsigli_spatial_2008}, and using neighbourhoods to create contingency tables \citep{ebert_fuzzy_2008, schwartz_comparison_2017}.

The FSS was originally proposed in \cite{roberts_assessing_2008} and \cite{roberts_scale-selective_2008}. Given $K$ forecasts and observations which cover a domain of size $N_x \times N_y$, the Fractions Skill Score is defined as:
\begin{align} 
\text{FSS}(n) &= 1 -  \frac{\sum_{k=1}^{K}\sum_{i=1}^{N_x} \sum_{j=1}^{N_y} ( f(n)_{ijk}  - o(n)_{ijk})^2}{\sum_{k=1}^{K}\sum_{i=1}^{N_x}\sum_{j=1}^{N_y} f(n)_{ijk}^2 + o(n)_{ijk}^2 } \label{eq:fss_main_sub}\\
&\equiv \frac{\sum_{k=1}^{K}\sum_{i=1}^{N_x} \sum_{j=1}^{N_y} 2 f(n)_{ijk} o(n)_{ijk}}{\sum_{k=1}^{K}\sum_{i=1}^{N_x}\sum_{j=1}^{N_y} f(n)_{ijk}^2 + o(n)_{ijk}^2 } \label{eq:fss_main}
\end{align}
where $f(n)_{ijk}$ ($o(n)_{ijk}$) is the fraction of forecast (observed) grid squares in the $k^{\text{th}}$ sample image that exceed an event threshold within a square window of width $2n +1$ centred at grid square $i,j$, where $ 0  \leq n \leq \max(N_x,N_y)$. Averaging over samples is typically performed separately for numerator and denominator before combining rather than taking an average of FSS values from different samples, since this reduces the possibility of comparing completely dry forecasts and observations, which result in an undefined score~\citep{mittermaier_meta_2021}. Other variants exist whereby neighbourhoods are constructed in time instead of space~\citep{woodhams_what_2018} and using ensembles~\citep{duc_spatial-temporal_2013, necker_fractions_2023}. 

A key part of interpreting the FSS is deciding on what level the FSS must reach such that a forecast is of high enough quality; this is referred to as `useful skill' in \cite{roberts_assessing_2008} and \cite{roberts_scale-selective_2008}. In \cite{roberts_assessing_2008} a method to interpret the skill from the FSS is provided, such that a forecast has useful skill if the FSS value exceeds a reference score of $(1 + o(0))/2$, where $o(0)$ is the frequency with which the precipitation event is seen in the observations at the grid scale. The same reference score has also been proposed as a means to estimate the displacement of precipitation objects \citep{skok_analysis_2015, skok_estimating_2018}, discussed further in Sec.~\ref{sec:useful}.

Despite its widespread use, there are two key problems with evaluating forecast skill by comparing with the reference score of $(1 + o(0))/2$. First, it is known that forecasts that do not exceed this score can still have considerable skill~\citep{nachamkin_applying_2015, mittermaier_long-term_2013}. Secondly, this reference score is derived at the grid scale, using inconsistent forecast definitions in the numerator and denominator \citep{skok_analysis_2015}, such that it does not have a straightforward interpretation across all neighbourhood sizes. With this as motivation, we present a much more robust method to assess forecast skill from the FSS, by deriving a baseline FSS score for random forecasts. We demonstrate that this derivation aligns precisely with FSS results for actual random data, and that it considerably changes how forecast skill is interpreted from the FSS. 

This paper is laid out as follows: In Sec.~\ref{sec:decomp} we present a decomposition of the FSS in terms of summary statistics. In Sec.~\ref{sec:useful} we explore existing approaches to derive skill from the FSS, and present a new method based on comparison with the FSS of a random forecast. Concluding remarks are given in Sec.~\ref{sec:conc}.

\section{Decomposing the FSS}
\label{sec:decomp}

We begin by rewriting the FSS score in eq.~\eqref{eq:fss_main} in a novel way that reveals the underlying factors that drive the score, and makes constructing a reference score possible. We use the angle bracket notation $\savg{x}$ to indicate the sample mean calculated over all grid points. Explicitly, it is defined as:

\begin{align}
    \savg{x} := \frac{1}{KN_xN_y}\sum_{k=1}^{K}\sum_{i=1}^{N_x} \sum_{j=1}^{N_y} x_{ijk}
\end{align}

Using this notation, the FSS equations in \eqref{eq:fss_main_sub} and \eqref{eq:fss_main} can be written as: 

\begin{align}
    \label{eq:fss_savg}
    \text{FSS}(n) & = 1- \frac{\savg{(f(n) - o(n))^2}}{\savg{f(n)^2 +  o(n)^2}} = 1- \frac{\savg{f(n)^2} + \savg{o(n)^2} - 2\savg{o(n)f(n)}}{\savg{f(n)^2} +  \savg{o(n)^2}} \\
    &=\frac{2\savg{o(n)f(n)}}{\savg{f(n)^2} +  \savg{o(n)^2}}  \label{eq:fss_main_2}
\end{align}
where $\savg{o(n)}$, $\savg{f(n)}$ are the sample neighbourhood frequency for observations and forecast respectively, calculated over all square neighbourhoods of width $2n +1$. 
We define $s_{o,n}, s_{f,n}$ as the (uncorrected) sample standard deviations for observations and forecast:
\begin{align}
   s^2_{o,n} &:= \savg{\left(o(n) - \savg{o(n)}\right)^2} = \savg{o(n)^2} - \savg{o(n)}^2 \label{eq:obs_sd_def}\\
   s^2_{f,n} &:= \savg{\left(f(n) - \savg{f(n)} \right)^2} = \savg{f(n)^2} - \savg{f(n)}^2
\end{align}
Note that these are biased estimates of the true standard deviations, since we are dividing by $KN_xN_y$, rather than $(KN_xN_y -1)$ \citep{von2002statistical}. Here we choose to use the biased estimator since it ensures that all terms have consistent denominators, and we assume that the domain has $N_x>10, N_y>10$ such that the biased and unbiased estimates will be very similar.
$r_{n}$ is defined as the sample Pearson correlation coefficient between the forecast and observed fractions:
\begin{align}
    r_{n} &:= \frac{1}{s_{f,n} s_{o,n}} \savg{\left(o(n) - \savg{o(n)}\right)\left(f(n) - \savg{f(n)} \right)} \nonumber\\
    &= \frac{1}{s_{f,n} s_{o,n}} \left(\savg{o(n)f(n)} - \savg{o(n)} \, \savg{f(n)} \right) \label{eq:sample_corr}
\end{align}

With these definitions we are now in a position to express eq.~\eqref{eq:fss_main_2} in terms of the sample statistics. A rearrangement of eq.~\eqref{eq:sample_corr} gives an expression for the numerator term:
\begin{align}
     \savg{o(n)f(n)} &= s_{o,n}s_{f,n}r_{n} + \savg{o(n)} \, \savg{f(n)} \label{eq:numer_expr}
\end{align}
Rearranging eq.~\eqref{eq:obs_sd_def} gives:
\begin{align}
   \savg{o^2(n)} = s_{o,n}^2 + \savg{o(n)}^2
\end{align}
and similarly for $\savg{f^2(n)}$, so that the denominator can be written:
\begin{align}
    \savg{f^2(n)} + \savg{o^2(n)} = s_{o,n}^2 + \savg{o(n)}^2 + s_{f,n}^2 + \savg{f(n)}^2 \label{eq:denom_expr}
\end{align}
Inserting eqs.~\eqref{eq:denom_expr} and \eqref{eq:numer_expr} into eq.~\eqref{eq:fss_main_2}, we arrive at a decomposed version of the FSS:
\begin{align}\label{eq:fss_exp}
\text{FSS}(n) & = \frac{2\savg{o(n)} \, \savg{f(n)} + 2 s_{o,n} s_{f,n} r_{n}}{\savg{o(n)}^2 + \savg{f(n)}^2 + s^2_{o,n} + s^2_{f,n}}
\end{align}

Despite the simplicity of this derivation, this expression of the FSS has not to the authors' knowledge been shown in existing literature, although decompositions of the mean squared error in this way are common (e.g.~\citet{murphy_skill_1988}), and a similar decomposition is arrived at in the context of the Intensity Scale Skill Score in \citet{casati_scale-separation_2023}. If we limit to the case where data at the grid scale have no spatial correlations, then $f(n)_{ijk}, o(n)_{ijk}$ are independent and Binomially distributed, and we recover the results in~\citet{skok_analysis_2016}. 

In order to show the explicit properties of the neighbourhood terms, we can arrive at expressions for $\savg{o(n)}, \savg{f(n)}, s_{f,n}, s_{o,n}$ in terms of quantities calculated at the grid scale and the spatial autocorrelations (see Appendix A). The effect of zero padding used to perform square convolutions at the edges (as is done for a standard implementation of the FSS \citep{pulkkinen_pysteps_2019}) makes the derivation of such expressions slightly more complicated. When using percentile thresholds to remove intensity biases, we observe that the neighbourhood frequency can still be reasonably different between forecast and observations when using zero padding, in contrast to using a scheme that pads with data from within the image, such as reflective padding. For this reason, and because it allows much simpler expressions for the neighbourhood mean and standard deviation later in this section, we calculate the FSS with reflective padding in this work. Another option is to not use padding, so that the number of grid cells to be compared shrinks as the neighbourhood size grows; we do not consider this in our work however a similar analysis would still apply with different definitions of how $\savg{o(n)}, \savg{f(n)}, s_{o,n}, s_{o,n}$ and $r_n$ are calculated. 

Derivations of neighbourhood mean and standard deviation under the assumption of reflective padding are given in Appendix A. The neighbourhood mean is equal to that at the grid scale, i.e.~$\savg{o(n)} = \savg{o(0)}$ and $\savg{f(n)}  = \savg{f(0)}$. The neighbourhood standard deviation $s_{o,n}$ can be written as:
\begin{align}
\label{eq:sigma_id}
s^2_{o,n} =  \frac{\savg{o(0)}(1-\savg{o(0)})}{(2n+1)^2}\left(
1 + \sum_{d=1}^{(2n+1)}\nu_o(d)\gamma_n(d) \right)
\end{align}
and similarly for $s_{f,n}$, where $\nu_o(d)$ ($\nu_f(d)$) is an estimate of the spatial autocorrelation between two grid squares a distance $d$ apart within the observations (forecasts), and $\gamma_n(d)$ accounts for the number of pairs of points within a neighbourhood that are separated by distance $d$. Thus $s_{o,n}$ and $s_{f,n}$ depend on $\savg{o(0)}$, the neighbourhood size, and the spatial autocorrelation.

\section{How to interpret the fSS}
\label{sec:useful}

 In this section we begin by summarising and clarifying previous results on how to interpret the FSS, before establishing a more robust method to assess forecast skill based on comparison to random forecasts.

\subsection{Summary of existing approaches}

We begin by summarising previously derived approaches for defining the no-skill to skill transition point from the FSS. In previous works, this has been defined as the point where the FSS for a forecast exceeds that of a simple reference score.

The first reference score is described in \cite{roberts_scale-selective_2008} as ``\emph{the FSS that would be obtained from a random forecast with the same fractional coverage over the domain as ... the base rate, [$\savg{o(0)}$] }". In other words, the score for a forecast that follows a Bernoulli distribution at the grid scale, with the Bernoulli probability set to $\savg{o(0)}$. This is given in \cite{roberts_scale-selective_2008} as:
\begin{align}
\label{eq:fss_rand_0}
    \text{FSS}_{\text{random}} = \savg{o(0)}
\end{align}
However, this reference score is only accurate for a neighbourhood size of 1 (i.e.~at the grid scale), and we shall show later in this section how it may be derived more rigourously. Because eq.~\eqref{eq:fss_rand_0} scales with $\savg{o(0)}$, it is typically too small to be informative and so does not appear to be used in the literature.

The most widely used reference score is defined as ``\emph{The FSS that would be obtained at the grid scale ... from a forecast with a fraction equal to [$\savg{o(0)}$] at every point}"~\citep{roberts_scale-selective_2008}, defined as:
\begin{align}\label{eq:fss_thresh}
\text{FSS}_{\text{uniform}} = \frac{1}{2} + \frac{\savg{o(0)}}{2}
\end{align}
 However, as noted in~\citet{skok_analysis_2015}, eq.~\eqref{eq:fss_thresh} does not result from the description given in~\citet{roberts_scale-selective_2008} and in fact results from setting $f(0)_{ijk}=\savg{o(0)}$ in the numerator and using a random binary forecast with mean $\savg{f(0)} = \savg{o(0)}$ in the denominator. We can verify this by inserting these definitions into eq.~\eqref{eq:fss_main_sub}:
\begin{align}
\text{FSS}_{\text{uniform}} &= 1 - \frac{\savg{(\savg{o(0)} - o(0))^2}}{\savg{f(0)^2 +  o(0)^2}}   = 1 - \frac{ s^2_{o,0} }{  \savg{f(0)^2} +  \savg{o(0)^2}}  \nonumber\\
&=1 - \frac{\savg{o(0)}(1-\savg{o(0)})}{2\savg{o(0)}} = \frac{1}{2} + \frac{\savg{o(0)}}{2}
\end{align}
where we have used the fact that $ \savg{o(0)^2} =  \savg{o(0)}$, and similarly $\savg{f(0)^2} =  \savg{f(0)} = \savg{o(0)}$, since the data is binary at the grid scale.

 Note that, if we take the same definitions for the $f(0)_{ijk}$ values but instead start from the rearranged form of the FSS in eq.~\eqref{eq:fss_main}, we arrive at a different result, since the numerator and denominator are not consistent with one another:
\begin{align}
\text{FSS}_{\text{uniform}} &= \frac{ 2 \savg{o(0)}  \savg{ o(0)}}{  \savg{f(0)^2} +  \savg{o(0)^2} }  = \frac{2 \savg{o(0)}^2}{2\savg{o(0)}} = \savg{o(0)}.
\end{align}

Since this reference score is derived using different forecast definitions on numerator and denominator, and is only derived at the grid scale, it has no obvious interpretation and does not necessarily scale properly with neighbourhood size. Previous work has also demonstrated that forecasts not exceeding this reference score can still have considerable skill \citep{nachamkin_applying_2015}. 

A derivation of a similar reference score is shown in \citet{skok_analysis_2016}, when the forecasts and observations fractions are uncorrelated and follow a Binomial distribution (or equivalently, forecast and observation events follow a Bernoulli distribution at the grid scale). Under these idealised assumptions, the FSS is shown to be equal to the reference score when the average number of rainy grid squares within the neighbourhood equals 1. Whilst there is a more solid mathematical derivation to this, it is not clear why this is a sensible reference score with which to assess the skill of a forecast. It is also only derived for Binomially distributed data, so does not hold for real observations which have spatial correlations.

The point at which the FSS reaches $\text{FSS}_{\text{uniform}}$ is also motivated as a means to estimate the displacement of forecast objects. Intuitively, increasing the FSS length scale reduces the effects of position errors in the forecast, and the point at which the FSS meets a critical point contains information about the displacement of precipitation objects. It can be shown that for idealised narrow vertical rain bands and distant sets of circular rainfall patterns~\citep{roberts_scale-selective_2008, skok_analysis_2015, skok_estimating_2018}:
\begin{align}
\label{eq:fss_disp}
\text{FSS}(n) = 1 - \frac{d}{2n+1}
\end{align}
where $d$ is the displacement between forecast and observation. This motivates the comparison between the FSS and $\text{FSS}_{\text{uniform}}$ as a means to estimate forecast displacement. Numerical investigations have also been performed on geometric shapes with displacement, rotation and distortion, and perturbations of a single real forecast using spatial shifts and changing the bias by a multiplicative or additive factor \citep{mittermaier_intercomparison_2010, skok_estimating_2018, ahijevych_application_2009}.~\citet{skok_estimating_2018} also examined this inferred displacement from real forecasts compared to reanalysis data, and found that the displacement inferred from the FSS appears to correlate well with the actual displacement.

\subsection{An improved method to interpret the FSS}

Having summarised previous results, we now present a more meaningful method to interpret FSS scores. We do this by comparing the FSS score to the score that would be achieved by a random forecast that follows a Bernoulli distribution at the grid scale, with the Bernoulli probability set to $\savg{o(0)}$. Forecasts that achieve a FSS score exceeding this baseline are then interpreted as having skill relative to that reference. This aligns with the standard concept of skill as defined in e.g.~\cite{wilks_forecast_2019}, and also appears to be the original intention in \cite{roberts_scale-selective_2008} and \cite{roberts_assessing_2008}, where they refer to `useful skill'.

Note the difference from the work in \cite{skok_analysis_2016}, in which both forecast and observation are assumed to follow a Bernoulli distribution, whereas here crucially only the forecast is. In \cite{skok_analysis_2016}, the authors use the simplified Bernoulli forecasts and observations to make the FSS mathematically tractable in order to study its properties, whereas here we are using the Bernoulli forecast as a baseline to compare against. Modelling the observations as following a Bernoulli distribution for this application is therefore inappropriate, as it would provide an unrealistic reference score.

We note that other definitions of `useful' are possible, and that in general these different definitions will give rise to different reference scores. This appears to be the case for using the FSS to estimate forecast displacement (as discussed in the previous subsection). Since there is empirical and theoretical evidence that the standard reference score in eq.~\eqref{eq:fss_thresh} can be used to measure forecast displacement, we regard this as a separate problem for which the standard reference score seems to function well.

We now show how skill relative to a random forecast can be derived for the FSS. Despite being named as a skill score, the FSS differs from other skill scores in that the reference forecast used is dependent on the forecast itself (and is in fact often unachievable by any forecast \citep{mittermaier_meta_2021}). This means that, unlike conventional skill scores, it is not straightforward to see whether or not a forecast has skill, which necessitates the following derivation. 

We start with eq.~\eqref{eq:fss_exp}, and note that a random forecast is not correlated with observations, so $r_{n}=0$. We take the Bernoulli forecast to have the same frequency at the grid scale, so $\savg{f(0)} = \savg{o(0)}$ and therefore, as discussed in Sec.~\ref{sec:decomp}, $\savg{f(n)} = \savg{o(n)}$. For the standard deviation term $s_{f,n}$, we use the expression for the neighbourhood standard deviation given by eq.~\eqref{eq:sigma_id} with $v(d)=0$ since the Bernoulli-distributed data is not spatially correlated. Substituting these into eq.~\eqref{eq:fss_exp} gives: 
\begin{align}
\label{eq:fss_rand}
\text{FSS}_{\text{random}}(n) &= \frac{2 \savg{o(n)}\savg{f(n)}}{\savg{o(n)}^2 + \savg{f(n)}^2 + s_{f,n}^2 + s_{o,n}^2} =
\frac{2 \savg{o(n)}^2}{2\savg{o(n)}^2 + s_{f,n}^2 + s_{o,n}^2} \nonumber \\
&= \frac{2 \savg{o(n)}^2}{2\savg{o(n)}^2 + \frac{1}{(2n+1)^2}\savg{o(0)}\left(1 -  \savg{o(0)} \right)+ s_{o,n}^2} 
\end{align}
We can see that for a neighbourhood width of 1 where $n=0$ and $s_{o,0}=\savg{o(0)}(1-\savg{o(0)})$, we recover the reference score in eq.~\eqref{eq:fss_rand_0} from \cite{roberts_scale-selective_2008} as expected:
\begin{align}
    \text{FSS}_{\text{random}}(0) &= \frac{2 \savg{o(0)}^2 }{2\savg{o(0)}^2 + \savg{o(0)}(1-\savg{o(0)})  + \savg{o(0)}(1-\savg{o(0)})} = \savg{o(0)}
\end{align}

In principle it is also possible to use this formula to represent an approximate FSS score for less simple reference forecasts, such as climatology or persistence. However, we would expect there to be non-zero correlation between observations and forecasts in such cases, which means an estimation of this correlation would be required. it may therefore be more insightful to simply calculate the FSS for a climatological forecast empircally rather than using a formula.

We now examine how comparing to the reference score in eq.~\eqref{eq:fss_rand} changes the interpretation of the FSS by plotting some illustrative examples on real data, chosen to highlight particularly interesting behaviours. For observations we use data collected by the Global Precipitation Measurement (GPM) satellites, processed using the Integrated Multi-satellitE Retrievals for GPM (IMERG) algorithm~\citep{huffman_integrated_2022}. For forecasts we use data from the European Centre for Medium-Range Weather Forecasts (ECMWF) Integrated Forecast System (IFS). Both datasets are regridded to $0.1^{\circ} \times 0.1^{\circ}$ resolution and hourly time steps, over the time period October 2018-June 2019 and over equatorial East Africa 12S-15N 25-51E; this is an area that has problems with extreme rainfall and drought, and in which standard rainfall parameterisation schemes typically struggle to perform well due to the dominance of convective rainfall~\citep{woodhams_what_2018}. For all examples, we use 90\textsuperscript{th} percentile thresholds to remove frequency biases, as is the recommended way to calculate the FSS in \cite{roberts_scale-selective_2008} and \cite{skok_estimating_2018}.

\begin{figure}[t]
  \centering
  \includegraphics[width=\textwidth]{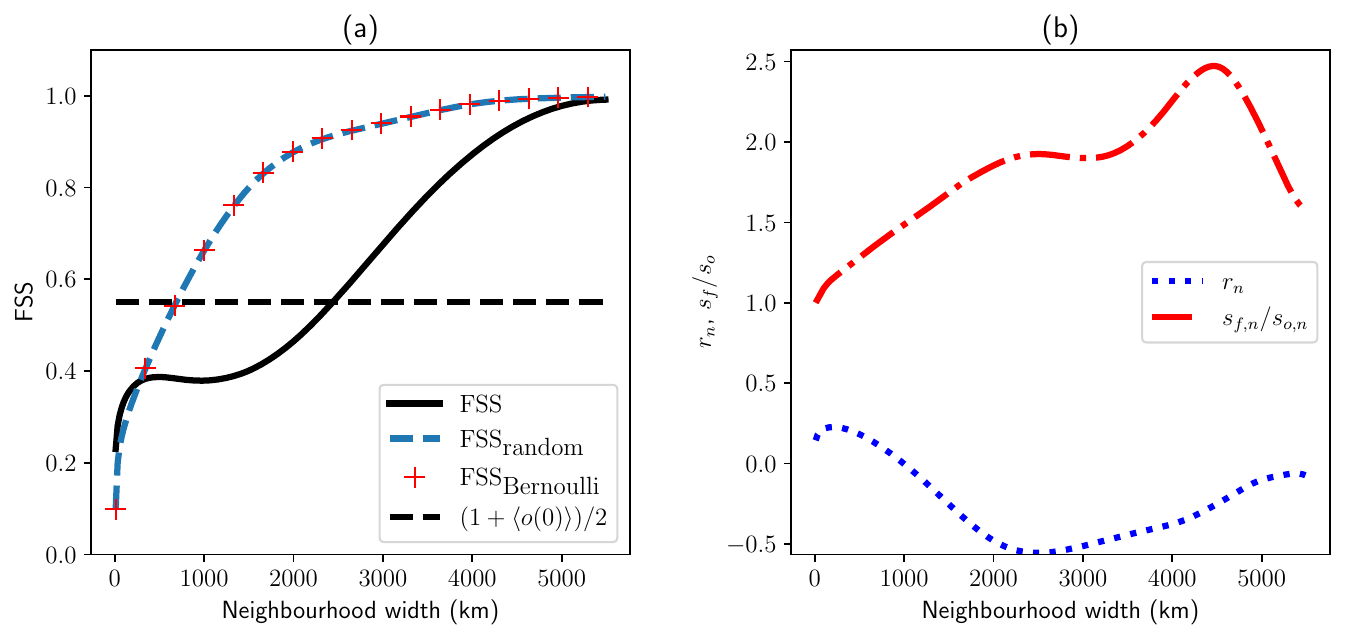}

\caption{Example FSS scores, using $90^{\text{th}}$ percentile thresholds, for 6hr accumulated rainfall between 18-24h on 15\textsuperscript{th} March 2019. (a) shows the FSS score (solid black line), the standard reference score for the FSS (black dashed line), the improved reference score based on random forecasts ($\text{FSS}_{\text{random}}$), and the FSS achieved from a Bernoulli forecast with the same frequency as the observations ($\text{FSS}_{\text{Bernoulli}}$). (b) shows the neighbourhood correlation $r_{n}$ and relative sizes of the neighbourhood quantities $s_{o,n}, s_{f,n}$ that contribute to the FSS score. Note that $\savg{f(n)}=\savg{o(n)}$ since we are using percentile thresholds.}
\label{fig:fss_example_1}
\end{figure}

\begin{figure}[t]
  \centering
  \includegraphics[width=\textwidth]{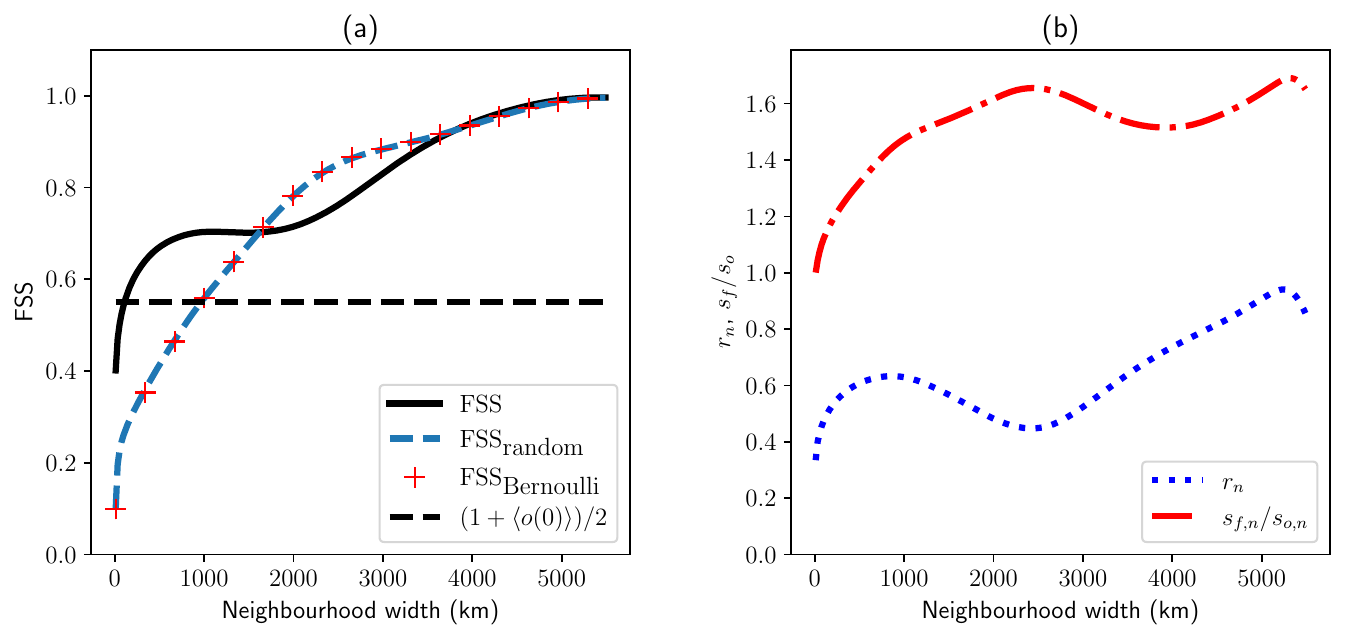}

\caption{As in Fig.~\ref{fig:fss_example_1} except for 6hr accumulated rainfall between 18-24h on 16\textsuperscript{th} March 2019. }
\label{fig:fss_example_3}
\end{figure}

\begin{figure}[t]
  \centering
  \includegraphics[width=\textwidth]{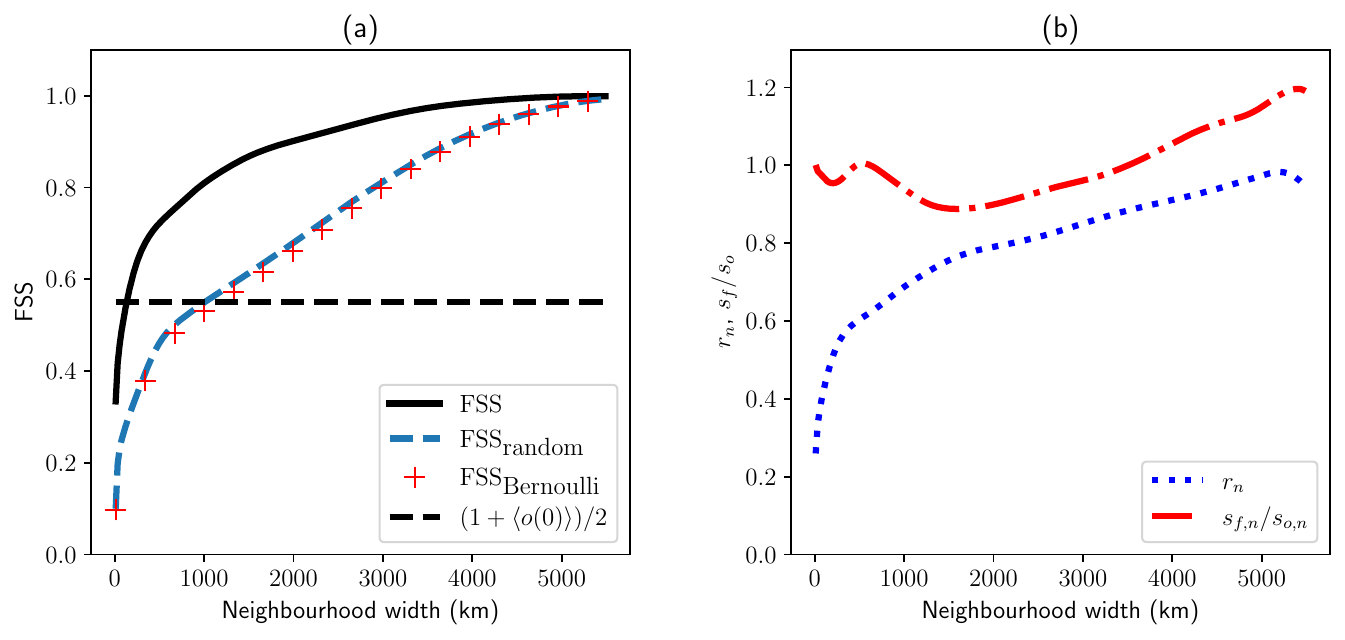}

\caption{As in Fig.~\ref{fig:fss_example_1} except for 6hr accumulated rainfall between 18-24h on 1\textsuperscript{st} March 2019. }
\label{fig:fss_example_2}
\end{figure}

\begin{figure}[t]
  \centering
  \includegraphics[width=\textwidth]{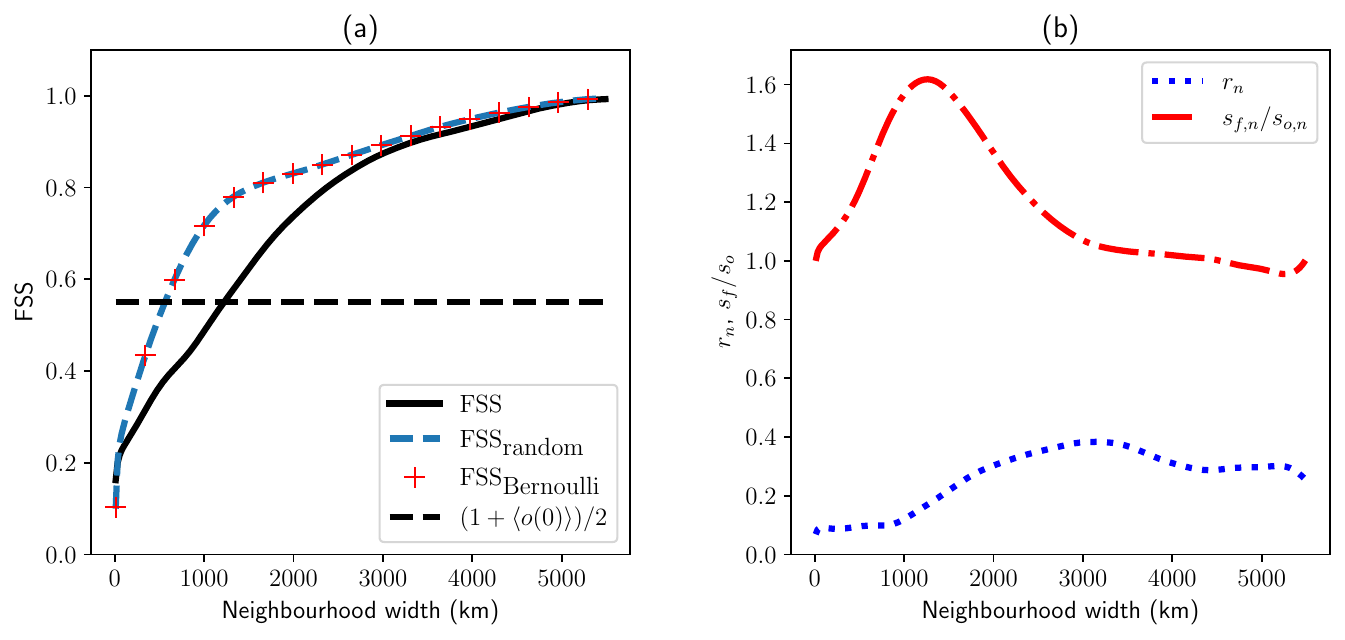}

\caption{As in Fig.~\ref{fig:fss_example_1} except for 6hr accumulated rainfall between 0-6h on 31\textsuperscript{st} May 2019. }
\label{fig:fss_example_4}
\end{figure}

In order to also validate what the true behaviour of a random forecast would be, we plot the results alongside the FSS for a random Bernoulli forecast having event frequency equal to the observed event frequency at the grid scale (labelled as $\text{FSS}_{\text{Bernoulli}}$ in the figures). Due to the domain size, the variability of the scores from these Bernoulli-distributed forecasts is small, hence only one sample is shown here. Alongside each plot of the FSS scores, we also show the values of $s_{f,n}/s_{o,n}$ and the neighbourhood correlation $r_{n}$, in order to illustrate the factors underpinning the scores (note that $\savg{f(n)}=\savg{o(n)}$ since we are using percentile thresholds). Our first observation of all of the examples in Figs.~\ref{fig:fss_example_1}-\ref{fig:fss_example_4}, is that the newly derived reference score is barely distinguishable to the FSS achieved from samples of Bernoulli forecasts, $\text{FSS}_{\text{Bernoulli}}$, confirming that the new reference score is a good approximation to that achieved by a random forecast. In contrast, the standard reference score bears no resemblance to it, including at the grid scale.

\begin{figure}[!ht]
    \centering
    \includegraphics[width=38pc]{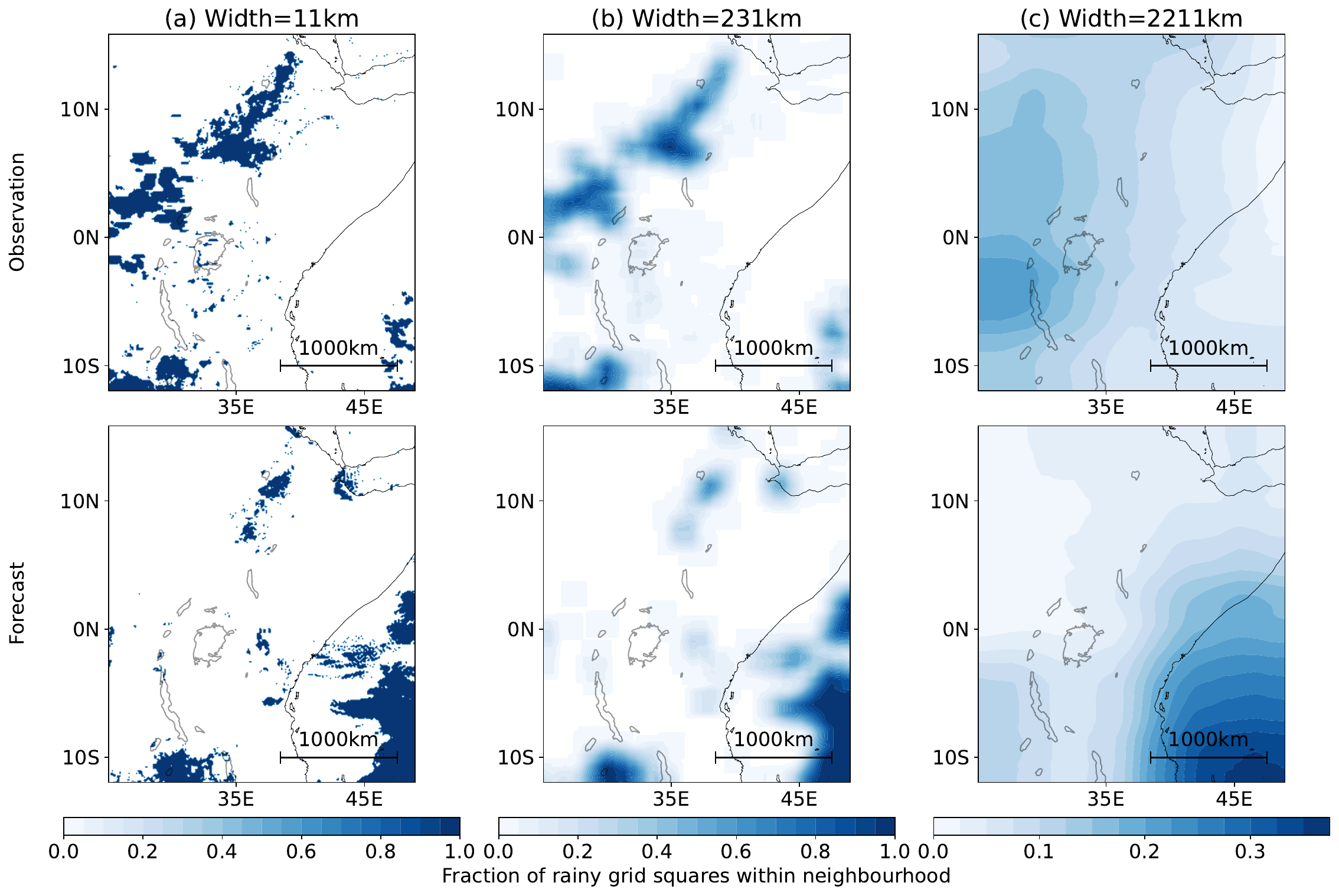}
    \caption{Images of the fraction of neighbouring grid squares at different neighbourhood widths for the first example, corresponding to the scores shown in Fig.~\ref{fig:fss_example_1}. Each column shows the result of converting observations (top row) and forecasts (bottom row) to a binary mask by applying a 90\textsuperscript{th} percentile threshold, and then calculating fractions of rainy pixels in a square neighbourhood around each pixel, with neighbourhood width given at the top of each column. Column (a) shows fractions with a neighbourhood width of 11km, column (b) shows fractions with a neighbourhood width of 231km, and column (c) shows fractions with a neighbourhood width of 2211km (around the point where the neighbourhood correlation is maximally negative).}
    \label{fig:blurred1}
\end{figure}

Maps of forecast and observation fractions for the first example, calculated over three different neighbourhoods, are shown in Fig.~\ref{fig:blurred1}. We can see that at a neighbourhood width of 231km (Fig.~\ref{fig:blurred1} (b)) the fields are slightly blurred but retain most of the structure, and at a much higher neighbourhood width of 2211km (201 grid points) the fields are very smooth, with highest fractions occurring in different parts of the domain.

The FSS curves for this example, shown in Fig.~\ref{fig:fss_example_1}, are striking in that the FSS curve meets the standard reference score (black dashed line) at a neighbourhood width of around 2000km, at which point there the neighbourhood correlation between the forecast and observations is  substantially negative. This can be seen in Fig.~\ref{fig:blurred1} (c), where the anti-correlation between the forecast and observation fractions is clear. In contrast, the newly derived reference score $\text{FSS}_{\text{random}}(n)$ is larger than the FSS curve within this range, and so correctly identifies this region of negative neighbourhood correlation as unskillful; only at low neighbourhoods (less than around 200km) is the forecast better than the random benchmark.

Similar although not quite as extreme behaviour is seen in Fig.~\ref{fig:fss_example_3}; the dip in neighbourhood correlation and increase in $s_{f,n}/s_{o,n}$ makes the FSS dip below $\text{FSS}_{\text{random}}$ at around 1000km, before exceeding it again at around 3000km. This and the previous example highlight that, contrary to the typical interpretation that there is a useful spatial scale beyond which the forecast is useful, there are in fact ranges of spatial scales for which the forecast has skill.

In Fig.~\ref{fig:fss_example_2} the FSS exceeds $\text{FSS}_{\text{random}}$ over all neighbourhood widths, and exceeds the standard reference score at around 100km. This highlights how the standard reference score can set much too high a bar at low neighbourhood sizes, and in some instances erroneously labels forecasts at the grid scale as unskillful. Notice that the uptick in the bias $s_{f,n}/s_{o,n}$ seen above a neighbourhood width of 4000km does not affect the score, since at this point $s_{o,n}$ and $s_{f,n}$ are much less than $\savg{o(n)}$ and $\savg{f(n)}$.

\begin{figure}[ht]
    \centering
    \includegraphics[width=38pc]{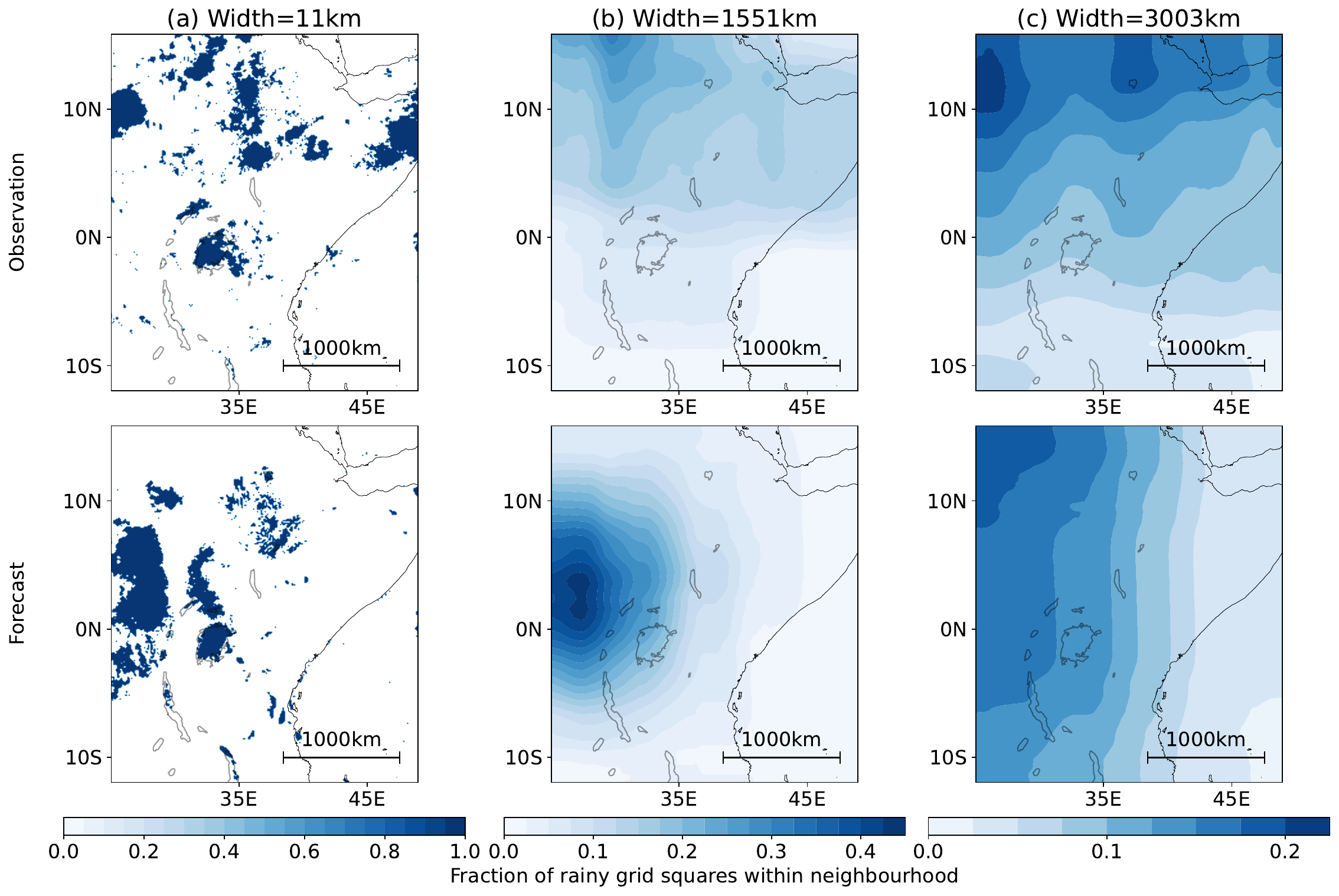}
    \caption{Images of the fraction of neighbouring grid squares at different neighbourhood widths, for the case in Fig.~\ref{fig:fss_example_4}. Each column shows the result of converting observations (top row) and forecasts (bottom row) to a binary mask by applying a 90\textsuperscript{th} percentile threshold, and then calculating fractions of rainy pixels in a square neighbourhood around each pixel, with neighbourhood width given at the top of each column. Column (a) shows fractions with a neighbourhood width of 11km, column (b) shows fractions with a neighbourhood width of 1551 (around the peak in $s_{f,n}/s_{o,n}$), and column (c) shows fractions with a neighbourhood width of 3003km.}
    \label{fig:blurred2}
\end{figure}

In contrast to the example in Fig.~\ref{fig:fss_example_2}, the example in Fig.~\ref{fig:fss_example_4} shows a case where the FSS does not exceed $\text{FSS}_{\text{random}}$ for any length scale, despite crossing the standard reference score line at a width of around 1500km. For this example, the bias in the neighbourhood standard deviation  $s_{f,n} / s_{o,n}$ rises as the neighbourhood correlation does, with a net effect of no skill. This highlights the trade-offs that are being made between different forecast errors. Further insight for this example can be obtained from the plots of fractions in Fig.~\ref{fig:blurred2}. From eq.~\eqref{eq:sigma_id}, we can see that any differences in $s_{f,n}$ and $s_{o,n}$ must be due to the spatial autocorrelation, since we are using percentile thresholds which remove frequency biases. This is indeed what is seen at a neighbourhood width of 1551km in Fig.~\ref{fig:blurred2}; the forecast fraction is more densely concentrated, and so has larger spatial autocorrelations at ranges up to about 1000km, whereas the observations show a more diffuse pattern with lower short range spatial correlations. Whilst the standard reference score would not make this region of low skill visible, the large gap between the calculated FSS values and $\text{FSS}_{\text{random}}$ highlights more clearly which neighbourhood lengths are problematic, in a way that also agrees with the underlying differences in $s_{f,n}$ and $s_{o,n}$.

To summarise, in this section we have presented a more rigourous derivation of a reference score for the FSS, such that if the FSS exceeds this score the forecast can be seen as superior to a random forecast with event frequency equal to that of the observations. In contrast to the existing reference score, which is derived at the grid scale only and uses inconsistent terms in numerator and denominator, this new reference score scales appropriately with the neighbourhood size, and is mathematically consistent. This is verified by comparing both reference scores to the average FSS score achieved for random Bernoulli forecasts, for which the new reference score is a precise match.

Through illustrative examples we have also demonstrated how this new reference score significantly changes the interpretation of FSS results. One particularly striking example is that the FSS can exceed the standard reference score whilst being substantially negatively correlated with observations, even when other neighbourhood biases are small. In contrast, the newly derived reference score correctly identifies this as a region of no skill. These examples also show that it is more accurate to say that there are ranges of spatial scales that are skillful, instead of the typical interpretation that there is a spatial scale beyond which the forecast is skillful.

\section{Discussion and Conclusions}
\label{sec:conc}

In this work we have provided a new method for interpreting skill from  the Fractions Skill Score (FSS), by deriving a new reference score corresponding to the score achieved by a random forecast; a score that exceeds this new reference score can be said to have skill relative to the random forecast. In contrast to the standard reference score, which is derived at the grid scale and has unclear meaning due to the inconsistent use of terms in the derivation, this reference score aligns precisely with the FSS achieved for actual random data, and has a clear interpretation. It also considerably alters how the FSS would be interpreted in many situations, and therefore presents a significant improvement to the insights that can be drawn from the FSS. One particularly interesting example shows that a forecast can exceed the standard reference score when the neighbourhood correlation between forecasts and observations is substantially negative. In contrast, the FSS for this situation does not exceed the newly derived reference score, demonstrating that interpreting results relative to this new reference score align more with our intuitions of skill. Therefore we recommend that FSS results should be assessed relative to the improved reference score presented in this work in place of the conventional approach, or else directly compared to other simple baselines, such as climatology or persistence.

We stress that this work focuses on the use of the FSS to assess the skill of a forecast, and not for other purposes such as estimating forecast displacement, as is done in e.g.~\cite{skok_estimating_2018}. Given the empirical and theoretical results that demonstrate how forecast displacement can be estimated from the FSS, the standard reference score of the FSS seems appropriate for these purposes.

\clearpage
\acknowledgments
The authors are grateful to Fenwick Cooper and Llorenç Lledó for comments on an earlier version of this work. David MacLeod regridded the IFS forecast data used to illustrate the results.

%
%
\datastatement
The Python code and data used to create the plots in this work can be found at \url{https://github.com/bobbyantonio/fractions_skill_score}.


\appendix[A]
\appendixtitle{Mean and variance of neighbourhood fractions}
\label{app:var_frac}

Here we derive expressions for the mean and variance of a fraction produced by a square convolution over binary data. Define $\mathcal{D}$ as the domain of grid squares over which the neighbourhood mean and standard deviation are to be calculated. Each location in this domain will be indexed by a single integer, to make the notation in this section easier to follow.

The fraction calculated over this neighbourhood at the location $i$, denoted $y_{i}(n)$, is given by the summation of values around the central point $i$ up to a distance of $n$ grid cells (i.e.~$y_i(n)$ is a placeholder for either the observed fraction $o_i(n)$ or the forecast fraction $f_i(n)$). We denote $W_n(i)$ as the set of all coordinate indexes that are within the neighbourhood of width $2n+1$ centred at point $i$. Then $y_{i}(n)$ is:
\begin{align}
y_{i}(n) = \frac{1}{(2n+1)^2}\sum_{j \in W_n(i)} x_{j}
\end{align}

\begin{figure}[t]
  \centering
  \includegraphics[width=\textwidth]{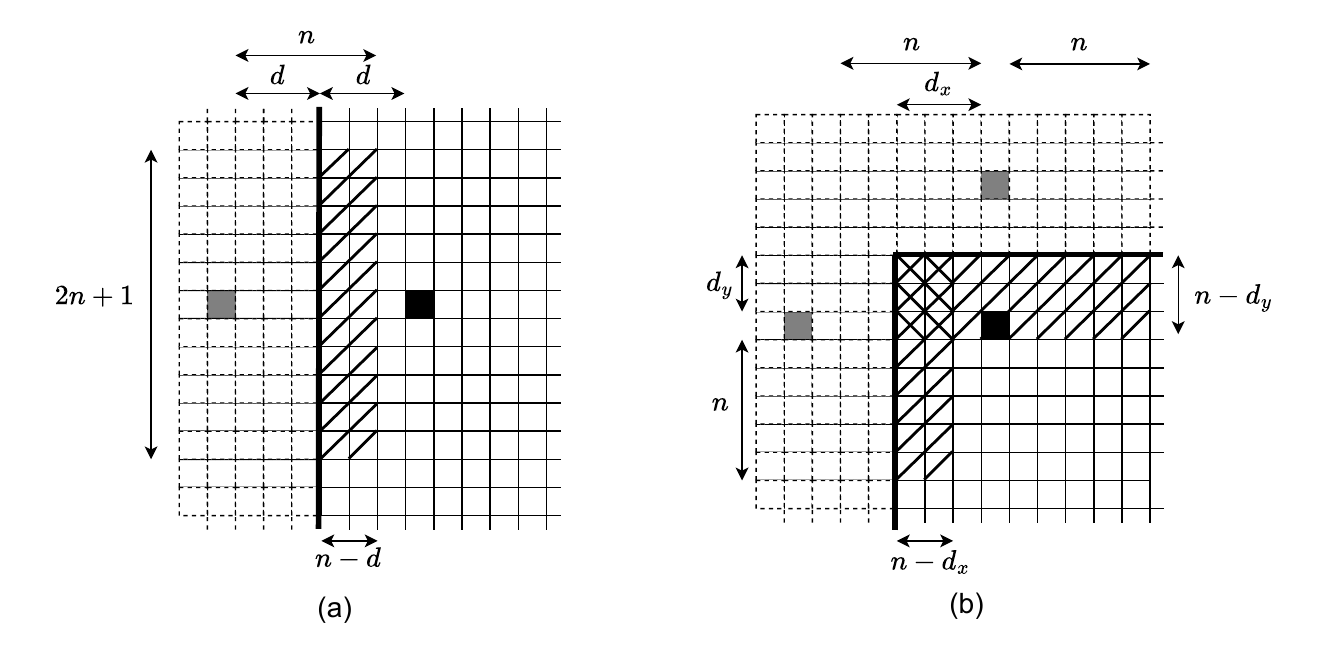}

\caption{Diagram illustrating how averaging over neighbourhoods behaves at the edges when reflective padding is used (a) for the case where a point is located a distance $d<n$ from an edge and (b) for the case where a point is located a distance $d_x<n$ from a vertical edge and $d_y<n$ from a horizontal edge. For both images, the point of interest is represented as a filled black square, the region of dashed lines represents the reflective padding, and the filled grey squares are where the original point is reflected to. Points that contain the reflected point once and twice are represented as single and double hatching, respectively.}
\label{fig:app_edge_case}
\end{figure}

The mean fraction is the average of $y_{i}$ over all sites; intuitively we can see that, since the averaging is a linear operation, the sample average $\savg{ y(n) }$ will be approximately equal to the sample average of the individual sites excluding padding, $\savg{x}$. The complicating factor is the padding used to compensate for the finite domain size; however it can be shown that with reflective padding, this relationship is in fact an equality. To show this, we first explicitly write out the sample average of $y(n)$:
\begin{align}
\label{eq:app_mu}
\savg{y(n)} &= \frac{1}{|\mathcal{D}|} \sum_{i \in \mathcal{D}} \frac{1}{(2n+1)^2}\sum_{j \in W_n(i)} x_{j} \nonumber\\
&=\frac{1}{|\mathcal{D}|}\frac{1}{(2n+1)^2} \sum_{i \in \mathcal{D}} \sum_{j : i \in W_n(j)} x_{i}
\end{align}
where in the last line, we have simply rearranged the summation to be in terms of a sum over all neighbourhoods that contain the point $i$ (which is only possible because we are performing a summation over all points in the domain). For a point that lies away from the edges, it is straightforward to see that this point will be contained in the neighbourhood of $(2n+1)^2$ other points. For points near the edges and corners however, this is not as obvious. Consider a point lying near an edge (as shown in Fig.~\ref{fig:app_edge_case} (a)); if the point is a distance $d <n$ from the edge, then without padding it is no longer contained within the neighbourhoods of $(2n+1)(n-d)$ points (i.e.~any points that lie within a distance of $n-d$ from the edge). With reflective padding however, this point is included in several neighbourhoods twice; the number of such neighbourhoods is equal to the number of points lying within a distance $n-d$ from the edge, which is also $(2n+1)(n-d)$. Thus each point lying along an edge is also contained within $(2n+1)^2$ neighbourhoods. 

The same can be seen for corner cases. Consider a point situated in a corner a distance $d_y < n$ from the top edge and $d_x <n$ from the side edge, as illustrated in Fig.~\ref{fig:app_edge_case} (b). Without reflective padding, it is only contained within $(n+d_x+1)(n + d_y+1)$ neighbourhoods. With the inclusion of reflective padding however, this site is included in several neighbourhoods multiple times; each point in the singly hatched area in Fig.~\ref{fig:app_edge_case} (b) includes point $i$ one additional time, whereas each point in the doubly hatched area includes point $i$ two additional times. The area of the single hatched areas plus twice the doubly hatched areas is just $(n-d_y)(n + d_x + 1) + (n-d_x)(n+d_y+1)$. This then brings the total number of neighbourhoods that $i$ is included in up to $(2n+1)^2$. 

Applying this to \eqref{eq:app_mu}, we therefore see that, with reflective padding 
\begin{align}
    \savg{y(n)} = \savg{y(0)} = \savg{x}
\end{align}

The (biased estimate of the) sample variance calculated over all fractions $y_{i}$, denoted $s_n^2$, can be written as: 
\begin{align}
s_n^2 &= \savg{y(n)^2} - \savg{y(n)}^2 \nonumber\\
&=  \frac{1}{|\mathcal{D}|} \sum_{i \in \mathcal{D}} \left(\frac{1}{(2n+1)^2}\sum_{j \in W_n(i)} x_{j} \right)^2  -  \savg{x}^2 \nonumber\\
&=  \frac{1}{|\mathcal{D}|} \sum_{i \in \mathcal{D}} \frac{1}{(2n+1)^4}\sum_{j \in W_n(i)} \sum_{k \in W_n(i)} ( x_{j}x_{k}   -  \savg{x}^2) \nonumber\\
&=  \frac{1}{|\mathcal{D}|(2n+1)^4}\sum_{i \in \mathcal{D}} \sum_{j \in W_n(i)} (x_{j}^2 - \savg{x}^2) + \frac{1}{|\mathcal{D}|(2n+1)^4}\sum_{i \in \mathcal{D}}\sum_{j \in W_n(i)} \sum_{\substack{k \in W_n(i) \\ k \neq j}} (x_{j}x_{k}  -  \savg{x}^2)\nonumber\\
&= \frac{1}{|\mathcal{D}|(2n+1)^4} \sum_{i \in \mathcal{D}} \sum_{j : i \in W_n(j)} (x_{i}^2 - \savg{x}^2) + \frac{1}{|\mathcal{D}|(2n+1)^4}\sum_{i \in \mathcal{D}} \sum_{j \in W_n(i)} \sum_{\substack{k \in W_n(i) \\ k \neq j}} (x_{j}x_{k}  -  \savg{x}^2)
\end{align}
where in the last line we have once again rearranged the sum to be in terms of the number of neighbourhoods containing point $i$, rather than the number of points in the neighbourhood of $i$. Using the same argument as above, the first term contains $(2n+1)^2$ copies of each summand, and so this simplifies to:
\begin{align}
    s_n^2&= \frac{1}{(2n+1)^2}(\savg{x^2} - \savg{x}^2) + \frac{1}{|\mathcal{D}|(2n+1)^4} \sum_{i \in \mathcal{D}} \sum_{j \in W_n(i)} \sum_{\substack{k \in W_n(i) \\ k \neq j}} (x_{j}x_{k}  -  \savg{x}^2) \nonumber\\
    &= \frac{\savg{y(0)}(1- \savg{y(0)}}{(2n+1)^2}\left[1 + \frac{1}{|\mathcal{D}|(2n+1)^2} \sum_{i \in \mathcal{D}} \sum_{j \in W_n(i)} \sum_{\substack{k \in W_n(i) \\ k \neq j}} \frac{(x_{j}x_{k}  -  \savg{x}^2)}{s_o^2} \right]
\end{align}
Where in the last line we have rewritten $\savg{x^2} - \savg{x}^2 = \savg{y(0)^2} - \savg{y(0)}^2 = \savg{y(0)}(1- \savg{y(0)}$ (because the data $y(0)$ is binary).

To simplify this further, we will group the terms inside the sum according to the $L_1$ distance (or taxicab norm) between them, where $L_1(i,j)$ denotes this distance (chosen since it is a natural metric for square neighbourhoods, but this could be substituted for other distance metrics with only slight modifications to the following derivation):
\begin{align}
\label{eq:app_stddev_1}
    s_n^2 &= \frac{\savg{y(0)}(1- \savg{y(0)}}{(2n+1)^2}\left[1 + \frac{1}{|\mathcal{D}|(2n+1)^2} \sum_{d=1}^{(2n+1)} \sum_{i \in \mathcal{D}} \sum_{j \in W_n(i)} \sum_{\substack{k \in W_n(i) \\ L_1(j, k) = d}}\frac{(x_{j}x_{k}  -  \savg{x}^2)}{s_o^2} \right]
\end{align}

Within a neighbourhood of size $(2n+1)\times (2n+1)$, we denote the number of points separated by a distance $d$ as $\gamma_n(d)$. With this notation, we define $v(d)$ as an estimate of the spatial autocorrelation for points a distance $d$ apart:
\begin{align}
\label{eq:spatial_autocorrelation}
    v(d) := \frac{1}{|\mathcal{D}|(2n+1)^2\gamma_n(d)} 
    \sum_{i \in \mathcal{D}} \sum_{j \in W_n(i)} \sum_{\substack{k \in W_n(i) \\ L_1(j, k) = d}}\frac{(x_{j}x_{k}  -  \savg{x}^2)}{s_o^2}
\end{align}
$v(d)$ will contain biases due to the reflective padding; near the edges, the correlation will be artificially inflated since any reflected points will be perfectly correlated with one other point in the neighbourhood. However, for our analysis, where the spatial autocorrelation term is only required to qualitatively understand what influences the value of $s_n^2$, this bias is acceptable.

Using the definition in \eqref{eq:spatial_autocorrelation} we can then rewrite \eqref{eq:app_stddev_1} as:
\begin{align}
    s_n^2 =\frac{\savg{y(0)}(1- \savg{y(0)}}{(2n+1)^2}\left[1 + \sum_{d=1}^{(2n+1)} \gamma_n(d) v(d)\right]
\end{align}

In the absence of spatial autocorrelation (i.e.~$v(d)=0$), $s_n^2$ is equal to the standard deviation for a Binomial distribution divided by $(2n+1)^2$ (since values are expressed as fractions).


\bibliographystyle{ametsocV6}
\bibliography{references}

\end{document}